**Type of Article:** original article

**Title:** Maize yield under a changing climate: the hidden role of vapor pressure deficit

**Authors list:** Jennifer Hsiao[1,*], Abigail L.S. Swann[1,2], Soo-Hyung Kim[3]

**Author Affiliations:**

[*] Corresponding author.

Email address: ach315@uw.edu

[1] Department of Biology, University of Washington, Seattle, Washington, USA 98195

[2] Department of Atmospheric Sciences, University of Washington, Seattle, Washington, USA 98195

[3] School of Environmental and Forest Sciences, College of the Environment, University of Washington, Seattle, Washington, USA 98195

**Abstract**

Temperatures over the next century are expected to rise to levels detrimental to crop growth and yield. As the atmosphere warms without additional water vapor input, vapor pressure deficit (VPD) increases as well. Increased temperatures and accompanied elevated VPD levels can both lead to negative impacts on crop yield. The independent importance of VPD, however, is often neglected or conflated with that from temperature due to a tight correlation between the two climate factors. We used a coupled process-based crop (MAIZSIM) and soil (2DSOIL) model to gain a mechanistic understanding of the independent roles temperature and VPD play in crop yield projections, as well as their interactions with rising $CO_2$ levels and changing precipitation patterns. We found that by separating out the VPD effect from rising temperatures, VPD increases had a greater negative impact on yield ($12.9 \pm 1.8\%$, increase in VPD associated with $2°C$ warming) compared to that from warming ($8.5 \pm 1.4\%$, the direct effect of $2°C$ warming). The negative impact of these two factors varied with precipitation levels and influenced yield through separate mechanisms. Warmer temperatures caused yield loss mainly through shortening the growing season, while elevated VPD increased water loss and triggered several water stress responses such as reduced photosynthetic rates, lowered leaf area development, and shortened growing season length. Elevated $CO_2$ concentrations partially alleviated yield loss under warming or increased VPD conditions through water savings, but the impact level varied with precipitation levels and was most pronounced under drier conditions. These results demonstrate the key role VPD plays in crop growth and yield, displaying a magnitude of impact comparative to temperature and $CO_2$. A mechanistic understanding of the function of VPD and its relation with other climate factors and management practices is critical to improving crop yield projections under a changing climate.

**Introduction**

Temperatures over the next century are expected to rise globally by 1.5-4˚C (IPCC 2013). Rising temperatures that exceed optimum for plant growth are considered detrimental for crop production (Sage and Kubien 2007, Battisti and Naylor 2009), leading many to question the sustainability of agriculture and food security under future warming (Lobell *et al* 2013, 2011b, Peng *et al* 2004, Asseng *et al* 2014). While much effort has been placed on understanding the negative yield impacts from temperature, the independent importance of vapor pressure deficit (VPD) is often neglected or conflated with temperature due to the tight correlation between the two climate factors.

VPD is an indicator for the dryness of the air, defined as the difference between the amount of moisture in the air and how much moisture the air can hold when it is saturated. Warmer air has a higher capacity to hold water, thus warming the atmosphere without additional moisture input leads to drying and higher VPD. VPD is not only an important atmospheric characteristic, but is also a key environmental factor that influences plant growth and development through mechanisms different from temperature (Eamus *et al* 2013, Day 2000, Shirke and Pathre 2004, Ray *et al* 2002, Sanginés de Cárcer *et al* 2018). Both warmer temperatures and accompanied elevated VPD levels can lead to negative impacts on crop yield, but through separate mechanisms. Warmer temperatures mainly affect plants through the temperature dependence of biochemical and developmental processes, influencing aspects of plant growth such as photosynthesis and developmental rate (Sage and Kubien 2007, Craufurd and Wheeler 2009). Elevated VPD, on the other hand, increases atmospheric water demand and plant water loss (Monteith 1995). This can trigger stomatal closure (Day 2000, Mott and Parkhurst 1991, Arve *et al* 2011) and several water stress responses such as lowered photosynthetic rates, reduced leaf area development, and an altered phenological timeline (Farooq *et al* 2009, Salah and Tardieu 1996, McMaster *et al* 2005).

Several aspects of VPD may have led to this discrepancy between the focus on temperature versus VPD within the literature. It is technically difficult to control water vapor levels independent of temperature under larger-scale experimental settings. Often, chamber, greenhouse or field studies aimed towards understanding temperature effects on crop growth also include effects of elevated VPD that are embedded within increased temperatures. Limitations also exist within empirical modeling approaches that rely on statistical relationships derived from crop yield observations and climate records over the past few decades. As formulated, these approaches typically cannot disentangle the inherent correlation between temperature and VPD. Temperature impacts on crop yield analyzed through these methods therefore include embedded VPD impacts as well (Lobell et al., 2011; Peng et al., 2004; Schlenker and Roberts, 2009).

A few empirical (Lobell *et al* 2014, Zhang *et al* 2017) and process-based modeling studies (Stöckle *et al* 2003, Lobell *et al* 2013) have recognized the importance of VPD effects on crop growth and yield, attributing mechanisms of yield loss under elevated VPD to decreased radiation use efficiency (Stockle and Kiniry 1990) and increased water demand and drought stress (Ort and Long 2014) in cropping systems. Studies have also shown VPD contributing to tree mortality in forested ecosystems (Breshears *et al* 2013, Eamus *et al* 2013) and altering water and carbon fluxes in various ecosystems (Novick *et al* 2016). While temperature and VPD are tightly correlated at present day, we do not expect the relationship to remain constant. Even under conditions of constant relative humidity, increasing temperature leads to a non-linear increase in VPD. Further, projections of future climate conditions suggest a decrease in relative humidity over land (Byrne and O'Gorman, 2016), implying even larger and more non-linear increases in VPD with warming. Understanding and quantifying the effects temperature and VPD independently have on crop yield is critical for future yield projections, especially when considering rising $CO_2$ concentrations.

Elevated $CO_2$ levels have the potential to alleviate part of the stress associated with hot and dry (elevated VPD) conditions either through a biochemical response in which more $CO_2$ boosts photosynthetic rate, or through a stomatal response in which stomatal closure under elevated $CO_2$ concentrations conserves water, improving plant water status and benefiting growth (Ghannoum *et al* 2000). The complex interaction between $CO_2$, VPD, temperature stresses, and water availability, however, lead to uncertainties when quantifying the positive $CO_2$ response. This is particularly true for $C_4$ crops, which often show little to no $CO_2$ responses in the absence of water stress (Chun *et al* 2011, Kim *et al* 2006, Leakey *et al* 2006, Manderscheid *et al* 2014). A better mechanistic understanding of the underlying processes through which each factor affects growth and yield, and how the effects of individual factors interact with one another as well as with background climate conditions would allow us to reduce uncertainty in yield responses to future climate conditions.

In this study, we applied a process-based modeling approach to disentangle the independent effects temperature and VPD have on crop yields. We focused on maize (Zea mays), the C4 cereal crop with the largest global production (Centro Internacional de Meioramiento de Maiz y Trigo CIMMYT and Pingali, 2001), by simulating yield through a coupled crop and soil modeling system developed and tested for maize (MAIZSIM-2DSOIL). Isolating the impacts from temperature and VPD allowed us to look into the interactions these two climate factors have with other aspects of a changing climate, such as rising $CO_2$ levels which benefit plant growth and changing precipitation patterns which could lead to water stress. We aim to 1) quantify the independent effects temperature and VPD have on maize yields, 2) gain a mechanistic understanding of the physiological processes that lead to their impacts on yield, and 3) identify their interaction with rising $CO_2$ levels and variation in precipitation patterns.

## 2. Materials and Methods

### 2.1 Crop simulation model

We selected a coupled crop (MAIZSIM) and soil (2DSOIL) model to represent the detailed processes that exist within the soil-plant-atmosphere continuum (Kim *et al* 2012, Yang *et al* 2009a, 2009b, Timlin *et al* 1996). We briefly describe here several key model components that were critical for this study. In this coupled model system, the model simulates maize growth and development as a function of light, temperature, precipitation, humidity, $CO_2$ concentrations, soil water content, soil properties, and nutrient levels. The model tracks key phenological stages to dynamically capture growth and physiological processes that are coupled with the surrounding environment while accounting for the effects of water and nutrient stress.

During the vegetative stage, non-linear temperature functions simulate emergence, maturation and senescence of individual leaves, while a logistic equation scaled by a temperature function is used to describe leaf expansion (Kim *et al* 2012). After transitioning into the reproductive stage, development is then determined by the ambient temperature and photoperiod (Kim *et al* 2012). The model describes gas exchange properties of developed leaves by coupling a leaf energy balance equation, a biochemical $C_4$ photosynthesis model (Von Caemmerer and Furbank 2003), and a stomatal conductance model (Ball *et al* 1987). Modifications in the stomatal conductance scheme were made to account for stomatal sensitivity to soil and plant water status. Gas exchange is calculated through iteratively solving for these three components (Kim and Lieth 2003), with carbon gained partitioned towards individual plant parts for growth, and water lost contributing to the overall water status within the soil and crop system.

The crop model is coupled with the 2DSOIL model that represents a two-dimensional soil domain capable of calculating heat and solute transportation (Timlin *et al* 1996). The coupled model dynamically solves for plant growth and development on an hourly basis, incorporating the real-time changes in the driving weather data. In our work, we utilized the mechanistic structure of this model to analyze the independent and interacting effects of temperature, VPD and $CO_2$ on crop growth, development and final yield.

*2.2 Model validation and application*

The MAIZSIM model has been previously validated for its representation of gas exchange processes (Yang *et al* 2009a), its leaf area simulation (Yang *et al* 2009b) under various water regimes, as well as its temperature response of leaf growth, development and biomass (Kim *et al* 2012). In addition to model validation with greenhouse datasets, MAIZSIM yield simulations have been tested against yield data collected from Free Air Carbon dioxide Enrichment experiments conducted at Thünen Institute in Braunschweig, Germany, to better understand yield simulations under elevated $CO_2$ conditions (Durand *et al* 2018). Yield simulations have also been validated with actual yield data in selected locations in France, the US, Brazil and Tanzania, along with 22 other maize simulation models (Bassu *et al* 2014), and applications of the model include estimating potential yield capacity in the U.S. Northeast Seaboard Region under current and projected future climate and land use scenarios (Resop *et al* 2016). These different angles of model validation and applications showed that MAIZSIM is capable of representing maize growth and yield under a range of environmental conditions, and its consideration of various plant responses to changes in temperature, $CO_2$ concentrations and different water stress levels make it suitable for exploring yield responses under a changing climate.

In our study, we compared model simulations of final yield with county-level yield data from the USDA National Agricultural Statistics Service's annual survey to validate the model's performance in the selected simulations (Fig. S1, United States Department of Agriculture, National Agricultural Statistics Service , NASS, http://www.nass.usda.gov/Quick_Stats). Since model simulations were not specifically calibrated for the cultivars, farming practices, and soil properties for each location, the main purpose of this comparison was not to evaluate whether the model accurately represented crop yield for each site, but rather to understand whether the model was able to pick up broad patterns of yield production. We up-scaled model yield outputs from single plant outputs (g/plant) to field-scaled units (tons/hectare) by assuming a uniformed planting density of 10 plants/m², and compared simulated model output of each site and year with actual yield data from the available yield data closest to each flux tower site (see site description below, Table S1, Fig. S2).

*2.3 Location selection and weather data*

The MAIZSIM model requires daily or hourly inputs of solar radiation, maximal air temperature, minimal air temperature, precipitation, relative humidity, and $CO_2$ concentrations for simulations. If a daily weather format is provided, the model runs an internal algorithm to interpolate the information into an hourly format. Detailed equations used for these estimations can be found in Timlin *et al*. (2002). We obtained all the information required for model simulation except for $CO_2$ concentrations through weather data archived within the AmeriFlux network (http://ameriflux.lbl.gov/). Based on availability of continuous daily weather data throughout the growing season, we selected four cropland flux tower sites in North America:

Iowa (US-Br1, Prueger and Parkin), Nebraska (US-Ne1, irrigated, Suyker; US-Ne3, rainfed, Suyker), Ohio (US-CRT, Chen), and Oklahoma (US-ARM, Biraud) (Fig. S1). We set $CO_2$ concentrations to a constant 400 ppm as a representation for current $CO_2$ levels. Sites were chosen to span background climate conditions with a range of mean annual temperatures and precipitation. Detailed information on individual sites can be found in Table S1. All available site data were used other than years with consecutive days of missing data (See below).

Ameriflux data are logged every 30 minutes, but gaps existed within the dataset. While MAIZSIM can also handle hourly and sub hourly weather input timesteps, we processed the 30-minute weather data into a daily input format to gap-fill short periods of missing weather data. To do so, we summed solar radiation over daylight hours, summed precipitation over a day, averaged relative humidity over a day, and selected daily maximal and minimal temperature. We defined a generic growing season starting from May until the end of October, and checked for missing meteorology data within this timeframe. Missing data that were less than half the length of a day were simply removed and daily averages were calculated with the remaining data. When missing data length was longer than half a day, we removed data from that entire day and gap filled it by interpolating from the day prior and after. When data had consecutive days of missing data, we removed that year of data from the analysis. We listed the years with available weather data used for each simulation site in Table S1.

*2.4 Idealized climate treatment*

We applied three main idealized climate treatments to the historical meteorological data to simulate the effect of *1)* warming through a 2˚C increase in temperature while holding VPD constant, *2)* atmospheric drying through an increase in VPD associated with a 2˚C warming while holding the temperatures constant, and *3)* increasing $CO_2$ levels from 400 to 800 ppm. We tested these climate treatments independently as well as in combination (Table 1). These climate changes follow the general magnitude of temperature change expected to accompany a doubling of atmospheric $CO_2$ concentration from 400ppm to 800ppm for mid-latitudes (Collins *et al* 2013). In addition, we imposed a 30% rainfall cut for the drought treatment to extend the precipitation range of our weather data to investigate the role of water stress. We included the simulated kernel yield for each treatment within Table 1 as well to provide an overall picture on how the different climate factors influences final yield.

Table 1. Treatment description and simulated kernel yield.

| Treatment | Description | Simulated Yield (g/plant) |
| --- | --- | --- |
| Control | Ameriflux weather data input. | $103.24 \pm 2.10$ |
| Elevated Temp | 2°C warming with VPD held constant. | $92.21 \pm 1.84$ |
| Elevated VPD | VPD increases that occur under a 2°C warming, but with temperature held constant. Average VPD increases across all simulation site and years are $1.19 \pm 0.66$ (kPa). | $88.71 \pm 1.85$ |
| Elevated $CO_2$ | Elevated $CO_2$ levels from 400ppm to 800ppm. | $117.98 \pm 2.22$ |
| Temp + VPD | 2°C warming plus increased VPD. | $77.33 \pm 1.84$ |
| Temp + $CO_2$ | 2°C warming plus elevated $CO_2$. | $106.15 \pm 1.97$ |
| VPD + $CO_2$ | Increased VPD plus elevated $CO_2$. | $109.66 \pm 2.20$ |
| All | 2°C warming, increased VPD, and elevated $CO_2$ combined. | $97.50 \pm 2.04$ |
| Drought | 30% rainfall cut. | $86.18 \pm 1.80$ |

We applied each climate treatment independently to the historical meteorology time series collected from the Ameriflux network archive. We followed the Clausius-Clapeyron equation to carry out our temperature and VPD treatments. For our simulations, we assumed that atmospheric water content (i.e. specific humidity) remains constant throughout warming, which results in a VPD increase with warming. The MAIZSIM model requires relative humidity as the humidity input, so we converted between VPD and relative humidity within our calculations. For the warming treatment, we applied a 2°C temperature increase throughout the growing season, and increased relative humidity to levels such that VPD would remain constant. Similarly, for the elevated VPD treatment, we calculated the increases in VPD that would have occurred under a 2˚C warming and converted it into relative humidity. We then applied the relative humidity change to our weather data while holding temperature constant. Due to the nonlinearity within the Clausius-Clapeyron equation, the magnitude of VPD change associated with a 2˚C warming would vary with the temperature and humidity levels at each simulation site-year; the averaged increases in VPD across all simulations were $1.19 \pm 0.66$ (kPa), with minimal increases as low as 0.17 and maximal increases up to 5.35 (kPa). Since these two treatments result from direct manipulation of historical meteorology data, they preserve the natural climate variability that existed within growing seasons and between planting years as well as the correlations between variables other than the one being manipulated. We included the equations used for these calculations within our supplementary information.

The stomatal conductance model incorporated in MAIZSIM follows the original Ball-Woodrow-Berry model (BWB model) and simulates stomatal closure under low relative humidity within the atmosphere (Ball *et al* 1987). While relative humidity and VPD both quantify water content in the atmosphere, they change at different rates as temperature increases, and plants respond more directly to VPD than to relative humidity (Mott and Parkhurst 1991), presenting a possible limitation for our study. It is worth noting that a recent study from Franks *et al.* (2017) demonstrated that with appropriate calibration, the BWB model captured a similar stomatal response within a the range of 0.5-2 kPa when compared to a BWB model variant with modifications for tracking the stomatal response to VPD (Medlyn *et al* 2011). The bulk of our meteorological conditions fall within this range.

*2.5 Model setup and simulation protocol*

We setup the model to represent a generic maize cultivar that requires 1600 growing degree days to reach maturity, and can develop a maximum of 15 juvenile leaves. We set a standard planting date of May-15 and planting density of 10 plants/m$^2$, and supplied a total of 200 kg-ha$^{-1}$ of Nitrogen throughout the simulation, with half supplied as base fertilizer prior to planting and the rest as top-dressing a month after planting. We divided the soil into four layers, and subscribed the top two soil layers to have a 0.65-0.28-0.06-0.01 sand-silt-clay-organic matter ratio, and the bottom two layers to have a 0.75-0.20-0.047-0.003 ratio.

We ran the model with the historical meteorology data for a control model simulation. We then forced the model with increased temperature, increased VPD, elevated $CO_2$, rainfall reduction treatments, and the combination of each to analyze the independent and interacting effects within and between each factor. Finally, we analyzed model outputs of growing season length, photosynthetic rate, total leaf area, predawn leaf water potential, stomatal conductance, and final yield.

*2.6 Analysis of model responses*

We calculated the percent impact each treatment (Temp, VPD and $CO_2$) and the combinations each had on yield (Table 1) by subtracting the simulated yield under climate treatments (*Yield$_{treatment}$*) from the simulated yield under the controlled climate (*Yield$_{control}$*), and further divided the output with *Yield$_{control}$* (Eqn. 1):

$$Yield\ Impact = \frac{(Yield_{treatment} - Yield_{control})}{Yield_{control}} * 100\% \dots\dots\dots\dots\dots\dots\dots\dots\dots\dots\dots (1)$$

Next, we identified three main factors that could influence crop yield: growing season length which we defined as time from planting to maturity (days), mean photosynthetic rate during the grain-filling phenological stage ($\mu$mol m$^{-2}$ s$^{-1}$), and maximal total leaf area throughout development ($cm^2$). We calculated the percent change in each of these growth factors when subjected to temperature, VPD, or $CO_2$ treatments as follows (Eqn. 2):

$$\Delta Growth\ Factor = \frac{(Growth\ Factor_{treatment} - Growth\ Factor_{control})}{(Growth\ Factor_{control})} * 100\% \dots\dots\dots\dots\dots\dots (2)$$

$GrowthFactor_{treatment}$ represents the simulated value of one of these factors of interest (growing season length, photosynthetic rate, or total leaf area) under climate treatments (i.e. Temp, VPD and $CO_2$). $GrowthFactor_{control}$ represents the simulated value of the same growth factor but under control climate conditions.

In addition, we averaged daily values of predawn leaf water potential (MPa) and stomatal conductance (mol $m^{-2}s^{-1}$) for each treatment throughout four developmental stages: emergence, tassel initiation, silking, and grain filling. Further, we utilized the natural precipitation variability within our meteorology data to analyze yield impacts (%) of temperature, VPD and $CO_2$ across precipitation levels. We classified growing season precipitation into four categories: high (> 600 mm), medium (500-600 mm), low (400-500 mm), and very low (< 400 mm) precipitation and calculated yield impacts (%) of each treatment within each precipitation bin. We calculated standard errors to represent variability between sites and years for all analyses.

## 3. Results

### 3.1 General agreement between simulated and actual yield

We compared crop yields simulated under the control weather input with actual county-level crop yield data collected from the years and proximate locations of which the weather data were collected (United States Department of Agriculture, National Agricultural Statistics Service, 2014). Figure S2 shows the ability of the coupled model to simulate actual crop yield across the four simulation sites (map of site location shown in Fig. S1). The model was able to represent yield variations across simulation sites and years due to the different weather inputs despite not being calibrated for specific cultivar and management practice for these sites. Model simulations represented a wider range of final yield values compared to observations, overestimating in locations and years with higher yields (Nebraska).

*3.2 Yield responses to temperature, VPD, and $CO_2$*

We summarized the simulated yield (g/plant) under different treatments across years and sites in Table 1. When compared to the control treatment, both warmer temperatures and higher VPD levels independently lowered yield by $8.5 \pm 1.4$ % and $12.9 \pm 1.8$ %, respectively, while imposing the two factors simultaneously lowered yield by $24.2 \pm 2.8$ % (Fig. 1a). On the other hand, elevated $CO_2$ concentrations increased crop yield by $17.2 \pm 3.5$ % (Fig. 1a). This boost in yield decreased when an elevated $CO_2$ level was accompanied by warmer temperatures or increased VPD, and the combination of all three climate effects (Temp, VPD, $CO_2$) partially canceled each other out, overall leading to a smaller reduction in yield by $-4 \pm 3.4$ % (Fig. 1a).

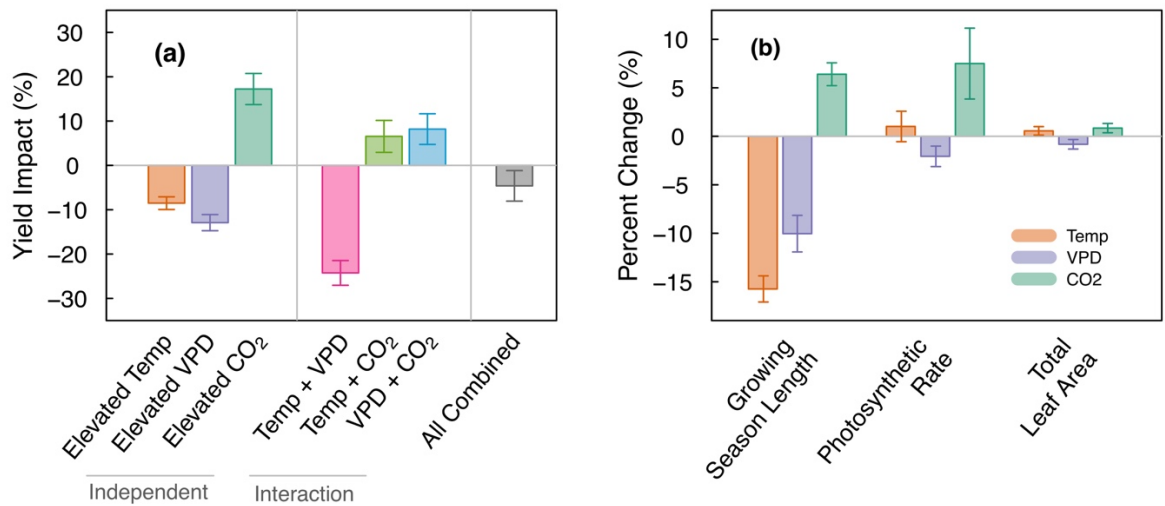

Figure 1. a) Percent yield impact from 2˚C warming, increased VPD that accompanies 2˚C warming, doubling of $CO_2$ levels from 400-800 ppm, and the combination of all three factors across all study

locations and years. b) The percent change in growth factors (growing season length, photosynthetic rate, total leaf area) under temperature, VPD and $CO_2$ treatments. Error bards denote standard error calculated across simulation sites and years.

*3.3 Mechanisms that contribute to changes in yield*

Shortened growing season length (days), lowered photosynthetic rate ($\mu mol m^{-2} s^{-1}$), as well as a decreased leaf area ($cm^2$) can all negatively affect final yield (Fig. 1b). When compared to the control simulation, the main cause of the negative yield impact from increased temperature (Fig. 1a, High Temp) was a shortened growing season (-15.7 ± 1.3 %, Fig. 1b), while effects from photosynthesis (1.0 ± 1.6 %) and leaf area development (0.6 ± 0.4 %) were neutral. Warmer temperatures did not significantly influence predawn leaf water potential (Fig. 2a) or stomatal conductance (Fig. 2c); the temperature effects illustrated in Figure 1b were mainly carried out through temperature alone. Note that the magnitude of these values should not be directly compared against one another, since a 5% shortening of growing season length will not necessarily have the same yield impact of a 5% decrease in photosynthetic rate. However, comparing these values between treatments (Temp, VPD, $CO_2$) can provide an overall picture on what growth factors are affected more by our imposed climate treatments.

Compared to rising temperatures, increased VPD levels had a greater negative impact on final yield (Fig. 1a, Elevated VPD). This is because a shortened growing season (-10.0 ± 1.9 %) was concurrent with lowered photosynthetic rates (-2.1 ± 1.1 %) and a slight decrease in leaf area development rate (-0.8 ± 0.5 %, Fig. 1b); these responses in photosynthetic rates and leaf area development to elevated VPD were in the opposite direction compared to the response under higher temperatures. Greater water loss under elevated VPD levels lowered predawn leaf water potential, leading to water stress signals that amplified later in the growing season (Fig. 2a).

Changes in water relations therefore played a critical role in the direction and magnitude of plant responses under an elevated VPD treatment (Fig. 1b).

In contrast to temperature and VPD effects, higher $CO_2$ concentrations benefited yield (Fig. 1a, High $CO_2$). Stomatal closure under elevated $CO_2$ concentrations (Fig. 2c) conserved water for use later in the growing season and improved water relations (Fig. 2a). This change in plant water relations directly counteracted negative VPD effects that stemmed from water stress responses (Fig. 2b), and partially alleviated yield through an increase in photosynthetic rates (7.5 $\pm$ 3.7 %), a prolonged growing season (6.4 $\pm$ 1.2 %), and a slight boost in leaf area development (0.8 $\pm$ 0.5 %, Fig. 1b).

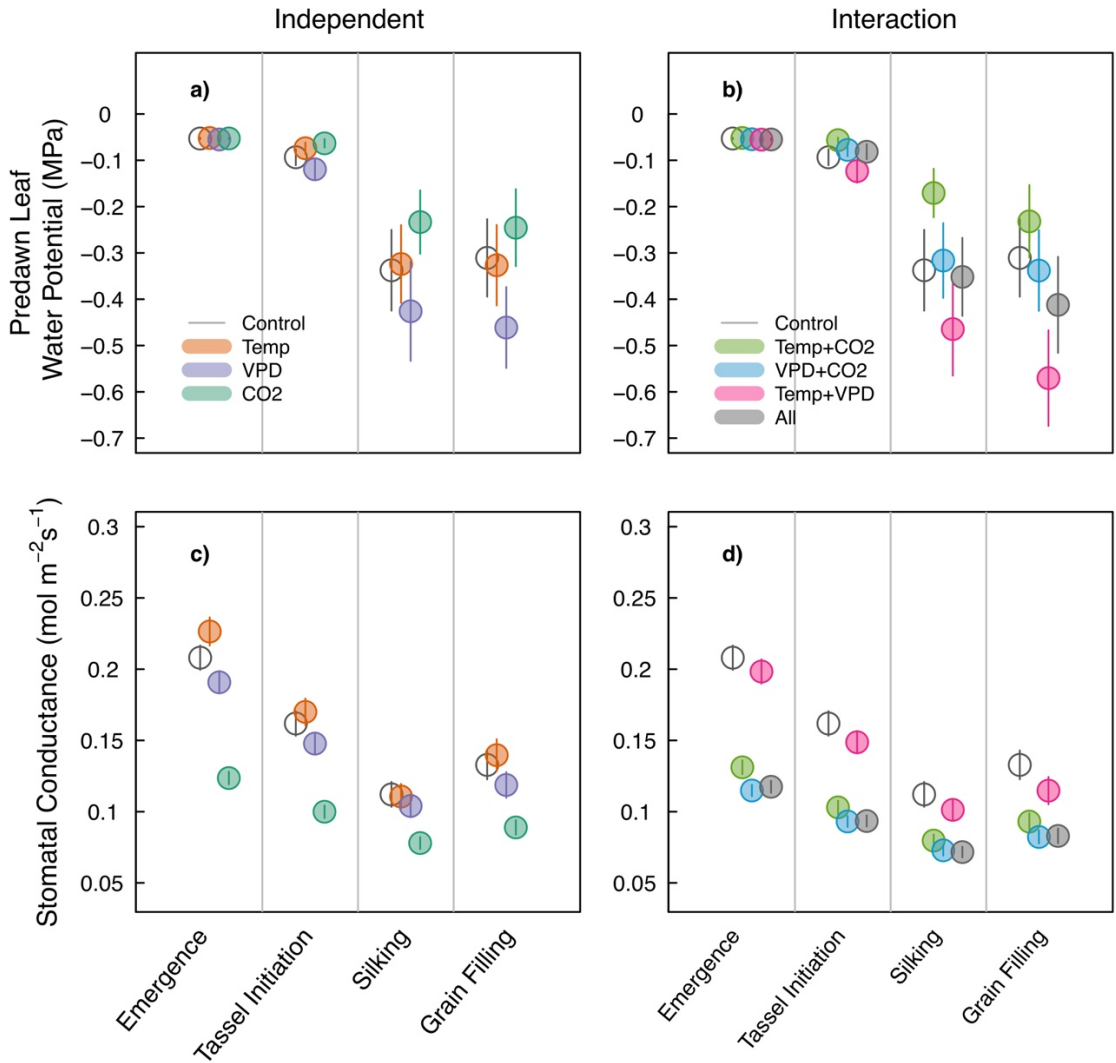

Figure 2. Average predawn leaf water potential (MPa) and stomatal conductance (molm$^{-2}$s$^{-1}$) across developmental stages of emergence, tassel initiation, silking, and grain filling for single climate treatments (a, c) and climate treatment combinations (b, d). Single climate treatments include Control (open circle), Temp (orange), VPD (purple), $CO_2$ (green), while climate treatment combinations include Temp+$CO_2$ (light green), VPD+$CO_2$ (blue), Temp+VPD (pink), and All (grey). Error bars denote variability between site and years.

### 3.4 The role of precipitation

For all treatments, the model generally simulated lower yield under lower precipitation. Over this natural range of precipitation variation present across the simulated sites and years, the direction of yield impact from temperature, VPD and $CO_2$ remained consistent, but the magnitude of impact each factor had on yield varied with precipitation levels (Fig. 3). Warmer temperature and higher VPD showed a greater negative relative yield impact in years and sites with medium-to-low precipitation levels, while the positive relative impacts from elevated $CO_2$ levels were most pronounced under drier conditions (Fig. 3d). Changes in simulated yield (Fig. 3a-c) correspond with the calculated percent impact on yield (Fig. 3d). The overall effect of temperature, VPD and $CO_2$ treatments combined led to a net negative effect on yield (Fig. 4a), with the greatest percent impact present in the medium precipitation range (Fig. 4b).

A 30% precipitation cut amplified the negative temperature and VPD impacts on yield (Fig. S3a & b) but diminished the positive $CO_2$ effects (Fig. S3c). These responses were most pronounced under site-years with lowest precipitation levels, and led to further yield loss (Fig. 4b). The negative yield impacts from temperature were greater than that from VPD under higher precipitation levels, but vice versa under very low precipitation levels (Fig. S3d).

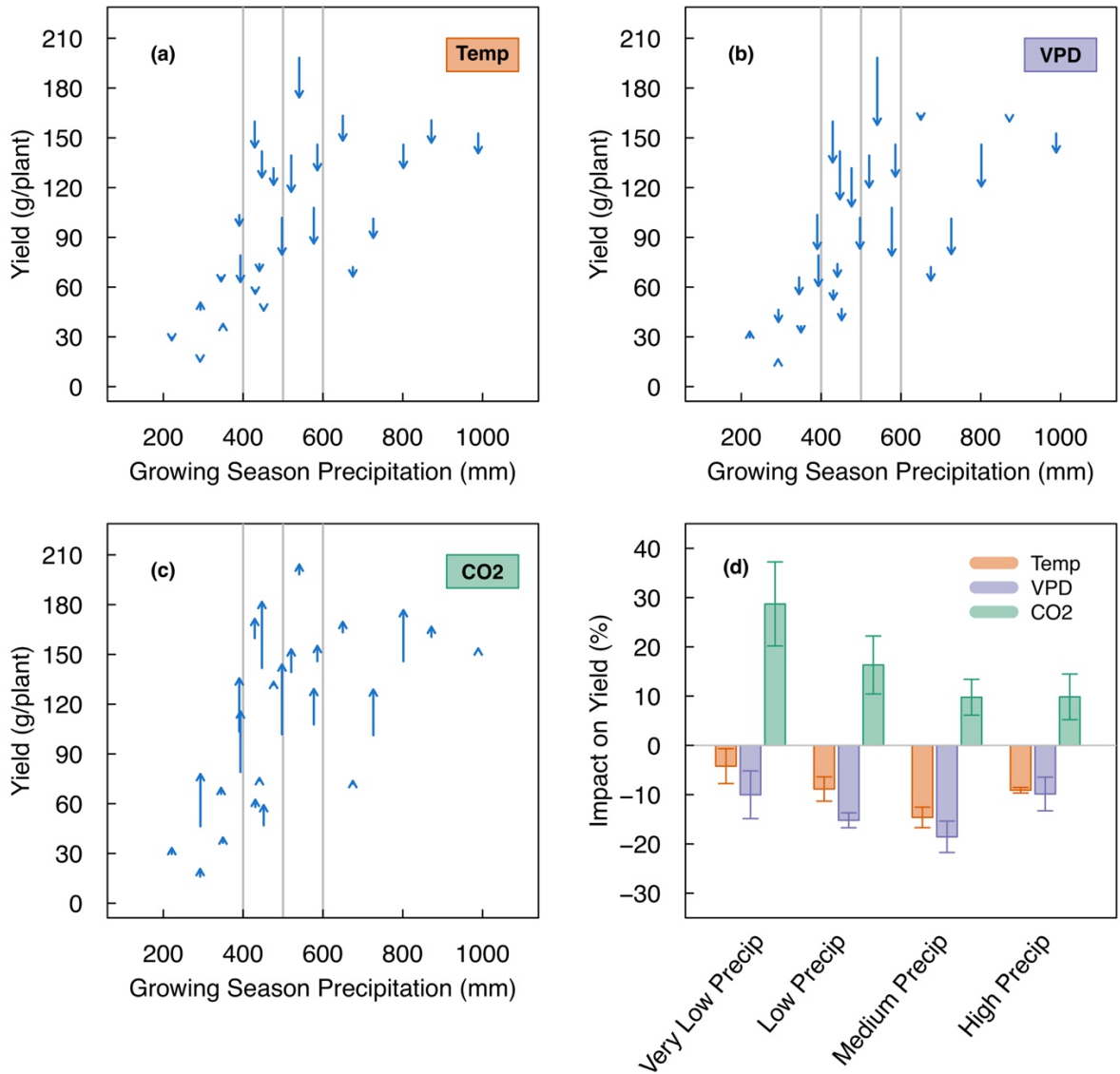

Figure 3. Direction and magnitude of yield change between treatment and control yield across precipitation range within simulated sites and years for a) temperature, b) VPD, and c) $CO_2$. The vertical grey lines categorize the precipitation range into very low (<400 mm growing season precipitation), low (400-500 mm), medium (500-600 mm), and high (>600 mm) precipitation levels. d) Percent yield impact from 2°C warming (orange), increased VPD that accompanies 2°C warming (purple), and doubling $CO_2$ levels from 400-800 ppm (green) under different precipitation ranges. Error bars denote standard error calculated across simulation sites and years.

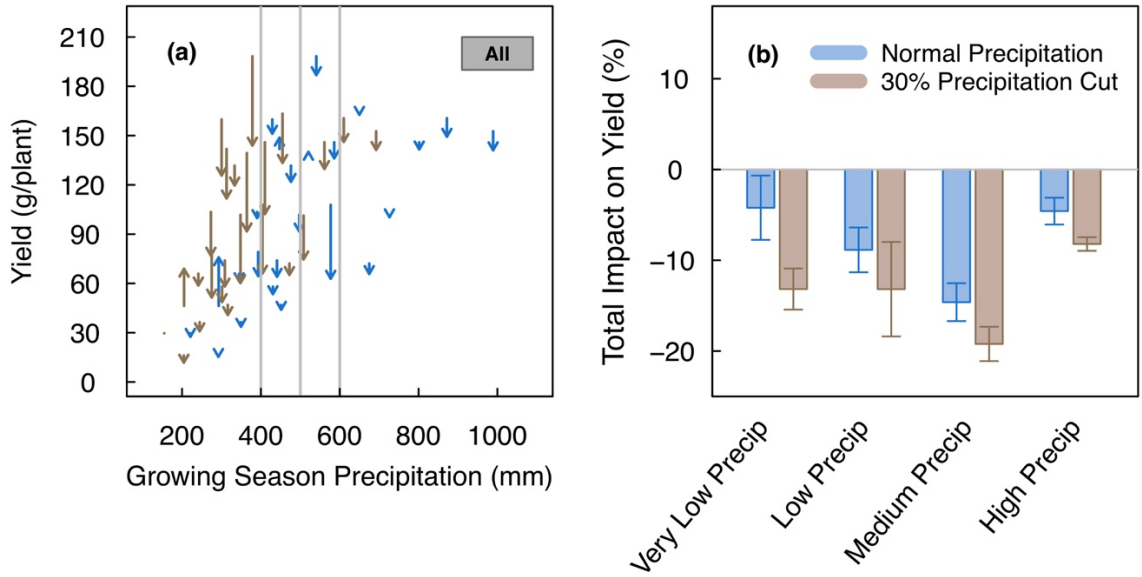

Figure 4. a) Direction and magnitude of yield change between the treatment (temperature, VPD and $CO_2$ combined) and controlled yield across precipitation range within simulated sites and years. The blue lines show results from the normal precipitation range, while the brown lines show results under a 30% rainfall cut. The vertical grey lines categorize the precipitation range into very low (<400 mm growing season precipitation), low (400-500 mm), medium (500-600 mm), and high (>600 mm) precipitation levels. b) Percent yield impact from temperature, VPD and $CO_2$ all combined across precipitation ranges for normal precipitation (blue) and 30% rainfall cut (brown).

## 4. Discussion

Both the 2 degrees of warming and VPD increase associated with that warming can independently lead to negative impacts on yield, with a larger yield decline due to the increased VPD than due to warming (Fig. 1a). Previous studies have largely focused on warming effects on yield loss (e.g. Lobell *et al* 2013, 2011b, 2008, Peng *et al* 2004, Asseng *et al* 2014, Battisti and Naylor 2009), but our results emphasize the importance of increased VPD levels that co-occur with warming. Elevated $CO_2$ levels, on the other hand, can partially alleviate yield loss from warming and increasing VPD, but at the levels we investigated (2 degrees warming, doubling of $CO_2$ from 400 to 800 ppm) the benefits from elevated $CO_2$ are not sufficient to completely compensate for the yield loss from warming and increased VPD.

*4.1 Effects of elevated VPD*

Under elevated VPD conditions, our simulation results showed lower photosynthetic rates, shortened growing season length, and a slight decrease in leaf area development (Fig. 1b). All of these factors contributed to a final loss in yield (Fig. 1a). These responses stemmed from increased water loss and water stress responses under a drier atmosphere (Fig. 2).

Various studies have demonstrated stomatal closure under dry atmospheric conditions in order to prevent excess water loss (Bunce 2006, Mott 2007, Monteith 1995), but despite stomatal closure, transpiration often still increases under the greater water vapor gradient between the leaf and the drier surrounding air (Monteith 1995). This phenomenon has also been demonstrated specifically in maize plants (Ray *et al* 2002). Greater water loss can increase the water stress levels experienced by plants, further influencing plant growth and yield through stomatal closure, interruption or reduction in leaf elongation and expansion (Çakir 2004, Nonami 1998, Tanguilig *et al* 1987, Salah and Tardieu 1996), or shortened phenology (McMaster *et al* 2005). We have accounted for these responses to water stress in our study, as the soil component within MAIZSIM tracks water status changes in the soil and the crop, quantifying it through soil and plant water potential (Fig. 2c). We find that predawn leaf water potential (and thus water status) continues to drop throughout the growing season under increased VPD (Fig. 2a) as stomatal closure is unable to fully compensate for drier air (Fig. 2b).Differences in the precipitation ranges experienced by individual site-years within treatment groups may have contributed to variations within simulated leaf water potential and stomatal conductance, but the overall pattern due to the climate treatments are still present.

The modified BWB model within MAIZSIM allows stomata to also respond to leaf water potential (Yang *et al* 2009a), in which water-stressed plants with low leaf water potential show stomatal closure (Fig. 2). While carbon limitation through stomatal closure has less of an influence on $C_4$ plants compared to $C_3$ (Ghannoum 2009), we still observed a lower photosynthetic rate (Fig. 1b) that accompanied stomatal closure under elevated VPD conditions (Fig. 2c), likely due to site-years under drier conditions. Lower leaf water potential in MAIZSIM also triggers a water stress response that reduces leaf expansion (Timlin *et al* 1996, Yang *et al* 2009b) and alters whole-plant carbon allocation such that more carbon is partitioned towards roots instead of aboveground plant parts, causing the reduction in total leaf area simulated (Fig. 1b). While phenology in MAIZISM does not directly respond to soil or leaf water potential, the changes in leaf elongation rate indirectly affected developmental time in each phenological stage, shortening the overall growing season length (Fig. 1b).

*4.2 Effects of elevated temperature*

Temperature, when separated from VPD, had little effect on plant water status (Fig. 2). Instead, the yield effects of temperature were directly through the temperature dependence of various plant processes. This aspect can easily be confounded when temperature and VPD are considered together. In contrast to VPD effects, increasing temperature showed a combination of positive and negative effects on phenological (growing season length), morphological (total leaf area) and physiological (photosynthesis) processes important for growth (Fig. 1b), which partially compensated each other such that the negative impact caused by warming was slightly lower than that caused by increased VPD (Fig. 1a).

The cumulative effect of warmer temperatures throughout the growing season showed a pronounced impact in accelerating development (Fig. 1b), moving the plants through phenological stages more quickly, reaching the reproductive stage and maturity earlier (Kim *et al* 2012), but the shortened developmental time also led to lower yield. Farming practices such as switching to more heat-tolerant cultivars and adjusting planting dates can modify the overall growing season length, mitigating yield loss under hot and high VPD conditions (Butler and Huybers 2012, Sacks and Kucharik 2011). We chose to set cultivar-related model parameters constant. This approach allowed us to look at yield impacts from temperature, VPD and $CO_2$ independent of other changing factors, but inevitably neglects the potential mitigating effects that can come from various farming and management practices.

Responses to warming in photosynthesis and leaf area development were lower in magnitude and varied in direction (Fig. 1b). Plants exhibit nonlinear temperature responses in various physiological processes including photosynthesis (Sage and Kubien 2007) and leaf growth (Kim *et al* 2007). As a $C_4$ crop, maize has a higher optimal temperature for many physiological processes compared to $C_3$ plants, and are generally more adapted to warmer climates (Collatz *et al* 1998). The direction and magnitude of temperature responses, therefore, depend greatly on whether the environmental temperature is greater or lower than the optimal temperature of each process (Kim *et al* 2007). Due to this nonlinearity in temperature responses, the warming treatment we imposed did not consistently lead to negative impacts on photosynthesis and leaf area development, but instead varied with the range of baseline temperatures found at each simulation site (Table S1). Cooler and wetter sites (i.e. Iowa, Nebraska) showed slightly positive temperature effects on photosynthetic rates ($4.37 \pm 1.27$, $4.49 \pm 2.91\%$, respectively). In contrary, the warmest and driest site (i.e. Oklahoma) showed negative effects ($-3.53 \pm 2.46\%$).

*4.3 Effects of elevated $CO_2$ and precipitation*

Decreased stomatal conductance under our elevated $CO_2$ simulation (Fig. 2c) led to water savings and increased predawn leaf water potential later in the growing season (Fig. 2a). While improved plant water status benefited yield, the overall effects were not sufficient to outweigh the negative effects from temperature and VPD (Fig. 1a). The magnitude of positive $CO_2$ impact on yield varied systematically with total growing season precipitation, with dry years and locations showing a greater mitigating effect from elevated $CO_2$ levels and vice versa (Fig. 3d). This response is commonly documented for $C_4$ plants, in which benefits of elevated $CO_2$ on crop growth are often only pronounced under water stressed conditions (Ghannoum 2009). Water savings have been more consistently documented in maize plants under elevated $CO_2$ conditions, either through lower evapotranspiration rates (Kim *et al* 2006) or water use (Chun *et al* 2011). On the other hand, $CO_2$ impacts on yield have been more variable and dependent on the level of water stress plants experience; chamber studies and Free-Air $CO_2$ Enrichment (FACE) experiments specific for maize showed little to no positive $CO_2$ effect on yield under well-watered conditions  (Chun *et al* 2011, Kim *et al* 2006, Leakey *et al* 2006), while yield boosts up to 40% were observed under droughted conditions (Manderscheid *et al* 2014).

The magnitude of positive $CO_2$ impact within our simulations range approximately from 10-30% (Fig. 3d). These values are comparable with FACE site studies that observed large positive yield boosts from elevated $CO_2$ concentrations (Manderscheid *et al.,* 2014). Durand *et al.* (2018) further tested how a collection of maize models represented the $CO_2$-water interacting effect demonstrated by Manderscheid *et al.* (2014) and revealed that despite a wide range of variability, models that specifically describe stomatal responses to $CO_2$ performed better in capturing the interacting effect between $CO_2$ and water status. MAIZSIM represents stomatal and photosynthetic response to $CO_2$, along with drought responses in leaf area development and carbon partitioning (Yang *et al*., 2009a, 2009b). It is also coupled with a soil module that explicitly tracks the movement of water down the soil column, and represents water potential both in the soil and in the crop (Timlin *et al*., 1996), making it suitable to explore the interacting effect between $CO_2$ and water status. In addition, our choice of soil composition resembles a well-drained soil, which can potentially lead to water stress under rain-fed conditions. This can partially explain the positive $CO_2$ effects that were still present under site-years with higher precipitation levels (Fig. 3d). Precipitation patterns throughout the growing season may also contribute to the positive $CO_2$ effects, even under site-years with higher total growing season precipitation since a high total growing season precipitation does not always correspond to optimal precipitation for maize growth. When looking into the growing season precipitation patterns, we noticed that site-years with more consistent rainfall showed less of a positive $CO_2$ responses (Fig. 3c, small arrows), while years with lower or more variable rainfall events, especially earlier on in development or during the grain-filling stage, resulted in greater positive $CO_2$ effects on yield (Fig. 3c, big arrows). This is likely due to the shallower rooting profile in younger plants within the model that make plants more susceptible to early growing season drought, and water-stress impacts on carbon allocation during the grain-filling stage, respectively.

The greatest absolute (Fig. 3a & b) and percentage (Fig. 3d) yield loss from warmer temperatures and elevated VPD occur under a mid-range precipitation level. This pattern suggests that sufficient rainfall partially alleviated the negative impact from elevated temperature and VPD, while droughted conditions served as an additional stressor on top of warm temperatures and elevated VPD (Farooq *et al* 2009), likely limiting the additional contribution of elevated temperature and VPD to yield loss. The interaction that temperature, VPD, and $CO_2$ effects exhibit with varying precipitation levels demonstrated how the relative contribution of each factor on final yield, as well as their combined effects, can shift with plant water status (Fig. 3d & 4b).

Plant water stress can arise from high atmospheric water demand (VPD), low belowground water supply (soil water potential), or a combination of the two (Novick *et al* 2016). Both processes can independently limit crop growth through their effects on plant water status (Salah and Tardieu 1996, Çakir 2004). While precipitation level is a quick indicator for water supply and potential drought, several other factors such as soil properties, root development, and rooting depth together determine final water availability.

*4.4 Future projections*

As temperatures are projected to continue rising over the coming decades, so are VPD levels (Collins *et al* 2013). Projected increases in VPD mainly follow the patterns of warming, as warmer air has the capacity to hold more water vapor, and thus the deficit from saturation increases with temperature. Even if relative humidity remained constant as temperatures increased, VPD would still increase due to the non-linear nature of the Clausius-Clapeyron relation. However, projections suggest decreasing relative humidity over land under a warmer climate (Byrne and O'Gorman 2016) which would lead to even larger increases in VPD. This poses a challenge in yield projection for models that do not consider temperature and VPD independently. Further, recent studies have also shown that stomatal closure under increasing $CO_2$ levels can limit water vapor input into the atmosphere, amplifying this trend of atmospheric drying due to vegetation and climate interactions (Berg *et al* 2016, Swann *et al* 2016). This decoupling between temperature and VPD is likely to lead to greater increases in VPD than predicted from temperature alone, emphasizing the importance of understanding the independent role of VPD in crop yield projections.

## 5. Conclusions

Our process-based modeling approach allowed us to quantify the direction and magnitude of the impact temperature, VPD and $CO_2$ independently had on final yield. While the absolute values shown in our results are specific to our model structure and setup, our results provided insight for the mechanisms behind these yield impacts: either through phenological (growing season length), morphological (leaf area development), or physiological (photosynthetic rate) processes. Our results illustrated that elevated VPD levels contributed greatly to the simulated yield loss under warming. This occurred through mechanisms linked to changes in water relations and were directly counteracted by water savings experienced under elevated $CO_2$ concentrations. These mechanisms were different from those that led to yield loss under warmer temperatures, which mainly acted through a shortened growing season. We also highlight the importance of water availability and its interaction with other climate properties (i.e. temperature, VPD, $CO_2$). While overall yield is reduced in lower precipitation years, the relative importance of each treatment factor on yield varies depending on the amount of precipitation. The relative yield boost from elevated $CO_2$ levels were larger in the lowest precipitation years and yield declines from warming and increased VPD were larger in the medium to low precipitation years. Such mechanistic and quantitative information on how different climate factors act independently and interactively to affect plant growth and yield can aid breeding programs for targeted breeding aimed towards climate change adaptation.

While climate plays a critical role for plant growth and production, management practices such as irrigation, fertilization, changes in planting dates or cultivars are also extremely important factors that affect the productivity within agricultural systems. Our work provides a starting point in examining the independent roles temperature and VPD play in crop yield projections, as well as their interactions with rising $CO_2$ concentrations and varying precipitation levels. Expanding this analysis to a broader geographic range with a wider representation of baseline climate conditions, while considering additional impacts from soil moisture, cultivar choices and farming practices, would provide further insight into the uncertainty in crop yield projections across agro-climate regions, and can provide guidance for acclimation and adaptation strategies moving forward under a changing climate.

**Acknowledgement**

We thank K. Yun for the technical support in setting up the MAIZSIM model. We also thank M.M. Laguë, M. Kovenock, G.R. Quetin, A.T. Lowe, and C. Crifo for feedback on the manuscript. JH and ALSS acknowledge support from the University of Washington Royalty Research Fund. ALSS acknowledges support from NSF award AGS-1553715 to the University of Washington, and S-HK acknowledges support from a Specific Cooperative Agreement (58-8042-6-097) with USDA-ARS and an Advanced Research Projects Agency - Energy award number DE-AR0000820 by the U.S. Department of Energy. Finally, funding for AmeriFlux data resources was provided by the U.S. Department of Energy's Office of Science.

**Supplementary Information**

*Disentangling Temperature and VPD*

We followed the Clausius-Clapeyron equation (Eqn. S1) to make independent adjustments to temperature and VPD. The equation uses the saturation vapor pressure ($E_{sref}$, 6.11mb) at a reference temperature ($T_{ref}$, 273.15 K), the vaporization latent heat (Lv, $2.5*10^6$ J/kg), and the gas constant (Rv, 461 J/K*kg) to calculate the saturated water vapor pressure ($E_s$) at air temperature T:

$$E_s = E_{sref} * e^{\left(\frac{Lv}{Rv} * \left(\frac{1}{T_{ref}} - \frac{1}{T}\right)\right)} \dots\dots\dots\dots\dots\dots\dots\dots\dots\dots\dots\dots\dots\dots\dots\dots\dots\dots\text{ (S1)}$$

With relative humidity (RH, %) available from our weather data, we can further calculate the actual water vapor pressure within the atmosphere (*E,* mb) through Equation S2, and VPD (mb) through Equation S3:

$$E = {(E_s * RH)}/{100} \dots\dots\dots\dots\dots\dots\dots\dots\dots\dots\dots\dots\dots\dots\dots\dots\dots\dots\dots\text{ (S2)}$$

$$VPD = E_s - E \dots\dots\dots\dots\dots\dots\dots\dots\dots\dots\dots\dots\dots\dots\dots\dots\dots\dots\dots\dots\text{ (S3)}$$

Under our 2°C elevated temperature treatment, $E_s$ simultaneously increases with temperature following equation S1. Normally, this would also cause an increase in VPD with an assumption that *E* remains constant throughout warming (Eqn. S3). To tease these two factors apart, we artificially increased RH levels such that VPD would remain constant (Eqn. S2, S3). Similarly, under our elevated VPD treatment, we calculated the VPD increase that would have occurred along with a 2°C warming (Eqn. S1, S3) and increased the VPD values within our weather data based on these calculations while holding temperature constant.

Supplementary Table 1. Description of Ameriflux tower data used for MAIZSIM weather input, and their mean growing season climate conditions and the standard deviation during simulation years.

| Location | Site Name | Latitude / Longitude | Site Description | Mean Growing Season Temp. (°C) | Total Growing Season Precip. (mm) | Mean Growing Season VPD (kPa) | Data Years Used | Citation |
|---|---|---|---|---|---|---|---|---|
| Iowa | US-Br1: Brooks Filed Site 10-Ames | 41.69 / -93.69 | Corn/soybean rotation cropland | 18.81 ± 6.03 | 730.76 ± 207.57 | 0.58 ± 0.31 | 2007-2008, 2010-2011 | Prueger and Parkin, doi:10.17190/AMF/1246038 |
| Nebraska | US-Ne3: Mead | 41.19 / -96.44 | Rain-fed maize-soybean rotation site | 19.42 ± 6.03 | 540.71 ± 152.53 | 0.72 ± 0.38 | 2004-2012 | Suyker, doi:10.17190/AMF/1246086 |
| Ohio | US-CRT: Curtice Walter-Berger Cropland | 41.63 / -83.35 | Soybean/winter wheat rotation* | 18.93 ± 5.82 | 466.43 ± 79.60 | 0.64 ± 0.38 | 2011-2013 | Chen, doi:10.17190/AMF/1246156 |
| Oklahoma | US-ARM: Southern Great Plains - Lamont | 36.60 / -97.49 | Winter wheat, corn, soybean, alfalfa | 23.53 ± 5.91 | 400.68 ± 127.48 | 1.18 ± 0.73 | 2003-2007, 2009-2011 | Biraud, doi:10.17190/AMF/1246027 |

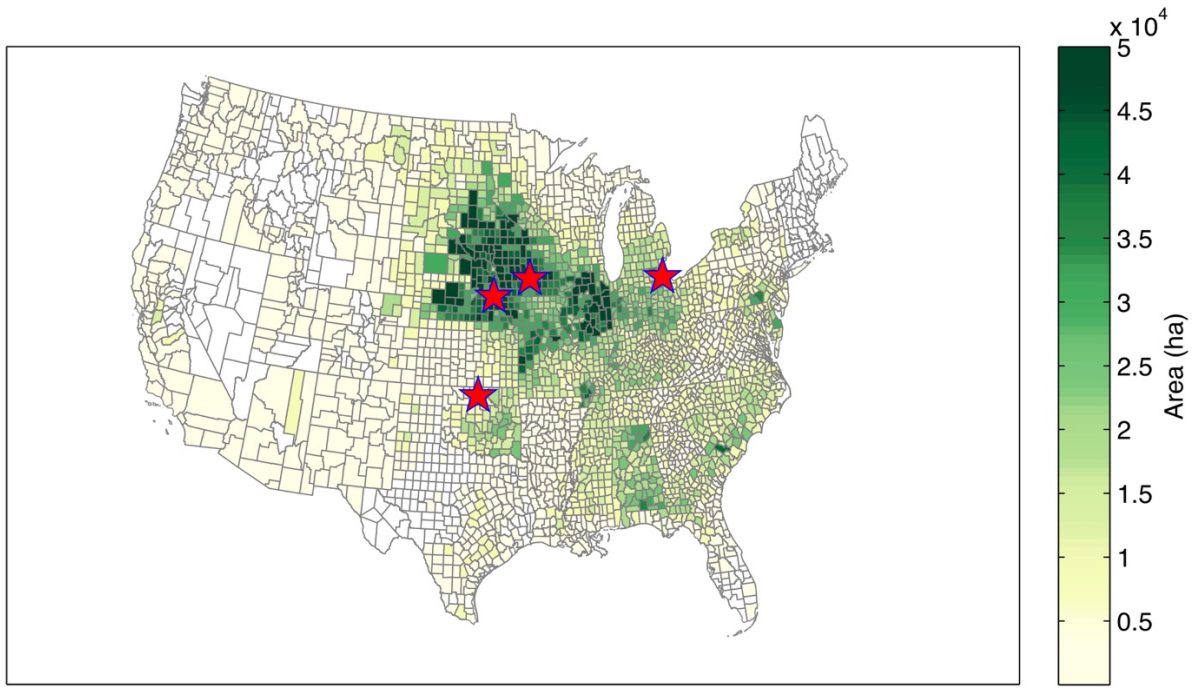

Supplementary Figure 1. Area planted in maize for all purpose production averaged across the most recent five years at the U.S. county-level, with the most recent year being 2014 (data from U.S. Department for Agriculture, National Agriculture Statistical Service). Red stars show the location of simulation sites.

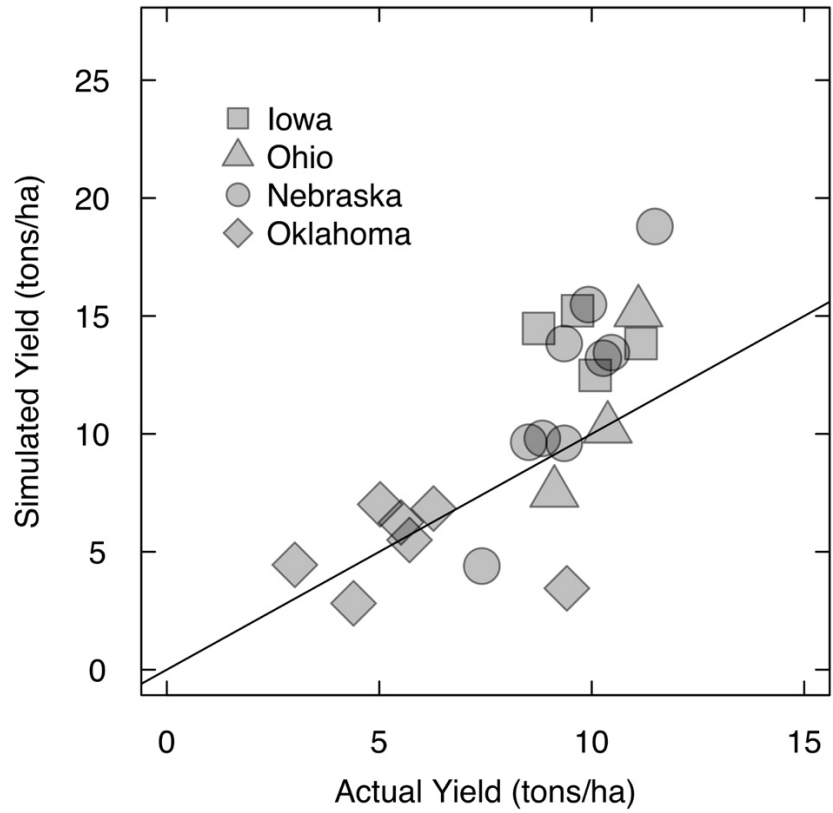

Supplementary Figure 2. Comparison between model simulation of final maize yield and actual yield observations across the study site and years (USDA NASS National Agricultural Statistics Service).

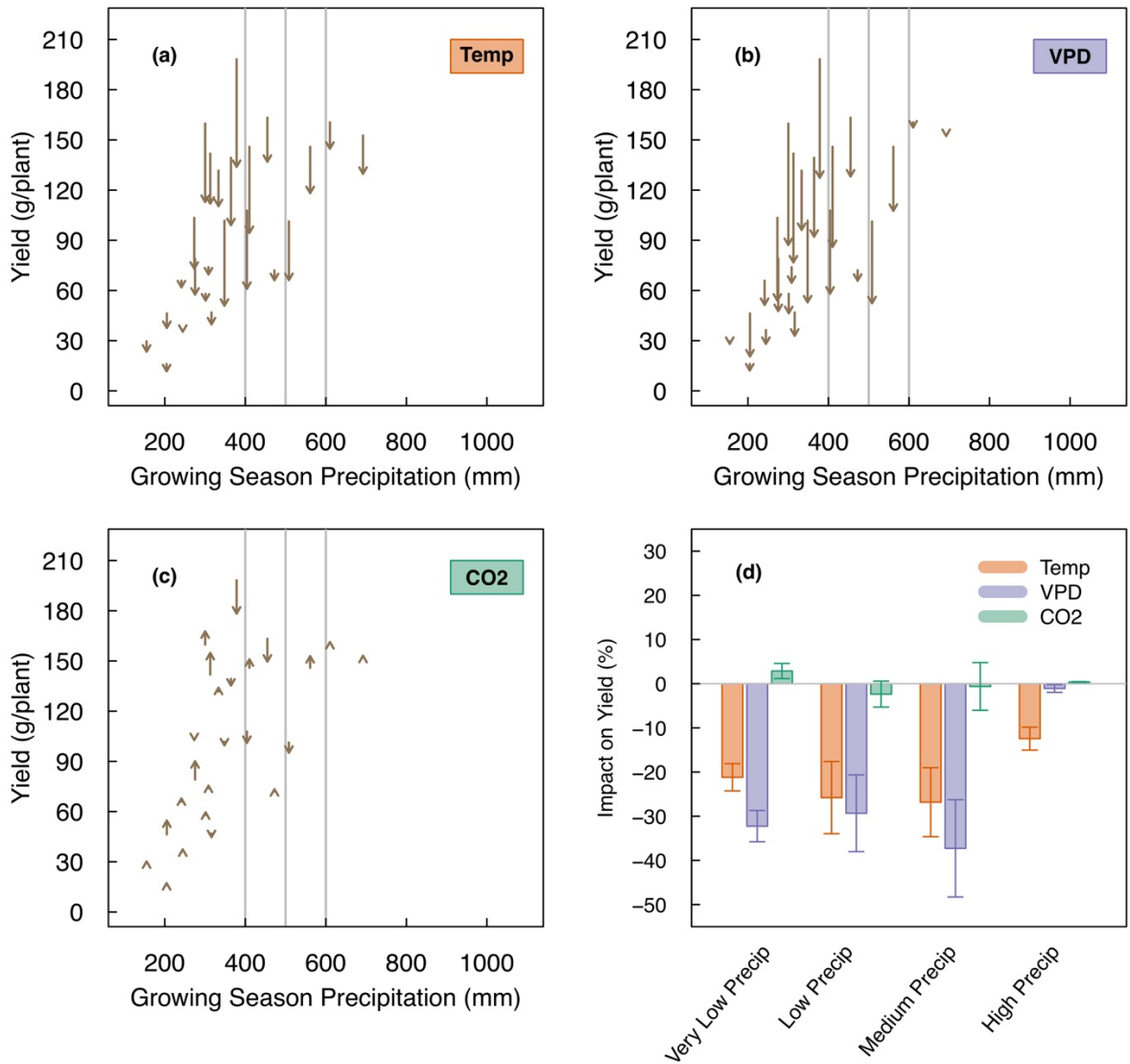

Supplementary Figure 3. Direction and magnitude of yield change between treatment and control yield under an additional 30% rainfall cut, across precipitation range within simulated sites and years for a) temperature, b) VPD, and c) $CO_2$. The vertical grey lines categorize the precipitation range into very low (<400 mm growing season precipitation), low (400-500 mm), medium (500-600 mm), and high (>600 mm) precipitation levels. d) Percent yield impact from 2°C warming (orange), increased VPD that accompanies 2°C warming (purple), and doubling $CO_2$ levels from 400-800 ppm (green) under different precipitation ranges. Error bars denote standard error calculated across simulation sites and years.